\documentclass[%
reprint, superscriptaddress,amsmath,amssymb,aps,
pre,prstper,floatfix,showpacs]{revtex4-1}

\usepackage[FIGTOPCAP,tight]{subfigure} 
\usepackage{graphicx}
\usepackage{dcolumn}
\usepackage{bm}

\usepackage{xcolor}

\newcommand{\pare}[1]{\left( #1 \right)}
\newcommand{\abs}[1]{\left\vert #1 \right\vert}
\newcommand{\cor}[1]{\left[ #1 \right]}

\newcommand{\ave}[1]{\left\langle #1 \right\rangle}

\begin{document}

\preprint{APS/123-QED}

\title{Generation of a tunable environment for electrical oscillator systems}

\author{R. de J. Le\'on-Montiel}
\affiliation{ICFO - Institut de Ciencies Fotoniques, Mediterranean
Technology Park, 08860 Castelldefels (Barcelona), Spain}
\author{J. Svozil\'{i}k}
\affiliation{ICFO - Institut de Ciencies Fotoniques, Mediterranean
Technology Park, 08860 Castelldefels (Barcelona), Spain}
\affiliation{RCPTM, Joint Laboratory of Optics PU and IP AS CR,
17. listopadu 12, 77146 Olomouc, Czech Republic}
\author{Juan P. Torres}
\affiliation{ICFO - Institut de Ciencies Fotoniques, Mediterranean
Technology Park, 08860 Castelldefels (Barcelona), Spain}
\affiliation{Department of Signal Theory and Communications,
Campus Nord D3, Universitat Politecnica de Catalunya, 08034
Barcelona, Spain}



\begin{abstract}
Many physical, chemical and biological systems can be modeled by
means of random-frequency harmonic oscillator systems. Even though
the noise-free evolution of harmonic oscillator systems can be
easily implemented, the way to experimentally introduce, and
control, noise effects due to a surrounding environment
remains a subject of lively interest. Here, we experimentally
demonstrate a setup that provides a unique tool to generate a
fully tunable environment for classical electrical oscillator
systems. We illustrate the operation of the setup by implementing
the case of a damped random-frequency harmonic oscillator. The
high degree of tunability and control of our scheme is demonstrated by gradually modifying the statistics of the
oscillator's frequency fluctuations. This tunable system can
readily be used to experimentally study interesting noise effects,
such as noise-induced transitions in systems driven by multiplicative noise, and noise-induced transport, a phenomenon that takes place in quantum and classical coupled oscillator networks.

\end{abstract}
\maketitle

\section{Introduction}

For many years, it has been known that fluctuations or noise can play an important role in various effects that take place in different physical, chemical and biological systems. Some examples of such effects are stochastic resonance \cite{hanggi_1998}, noise-induced transitions \cite{horsthemke_book,parrondo_1994}, and noise-induced transport, a phenomenon that has been observed in quantum \cite{aspuru_2009,plenio_2008} and classical \cite{hanggi_2009,roberto_2013} systems.

Because noise-induced effects are generally described by models where several, albeit reasonable, assumptions are made, an experimental confirmation of these surprising, sometimes counterintuitive, theoretical predictions is certainly most desirable. The verification of predicted noise effects is, in general, most easily achieved on simple experimental systems. As stated in Ref. \cite{horsthemke_book}, these systems should exhibit the following features: i) their time evolution should be well known for deterministic conditions, ii) their experimental design should not present great technical difficulties, and iii) variables of the system and the externally introduced noise should be easily controlled. In view of these points, we immediately realize that electrical oscillator circuits are the ideal choice. Indeed, the majority of the experimental studies on noise-induced phenomena has been carried out using electrical circuits \cite{kawakubo_1978,kabashima_1979,kabashima2_1979,berthet_2003}. Other systems where noise effects have been studied involve surface waves \cite{residori_2001}, spin waves in ferrites and antiferromagnets \cite{zautkin_1983}, and electroconvection in nematic liquid crystals \cite{john_1999}.

Most of the experiments mentioned above make use of systems driven by Gaussian white noise. However, it has been shown that systems driven by non-Gaussian noises might as well exhibit interesting features, such as shifts in the transition line for noise-induced transitions \cite{wio_2004}, enhancement of the signal-to-noise ratio in stochastic resonance \cite{fuentes_2001, castro_2001}, and enhancement of transport efficiency in Brownian motors \cite{bouzat_2004}. Therefore, in order to experimentally investigate new non-Guassian noise effects, one needs to design a system capable of producing various types of noise bearing different probability distributions.

In this paper, we introduce an experimental setup that performs as a tunable environment for classical electrical oscillators. We test our scheme by implementing the case of a damped random-frequency
harmonic oscillator. We have chosen this system because it represents a fundamental tool in statistical physics, which has been extensively used to describe a myriad of physical systems in different research fields \cite{gitterman_book,graham1982,ishimaru1999,turelly1977,takayasu1997}. The tunability of our system is demonstrated by gradually modifying the statistics of frequency
fluctuations, which is managed by properly controlling the mean and variance of the oscillator's frequency distribution. This is particularly relevant, because it implies that the system introduces directly fluctuations in the frequency of the signal, which contrasts with previous experimental studies, where fluctuations in the
amplitude, rather than frequency, are introduced in the system
\cite{residori_2001,berthet_2003}.

Because of its high degree of tunability and control, this setup
can readily be used to experimentally observe effects of
Gaussian \cite{berthet_2003,gitterman2013} and non-Gaussian \cite{wio_2004} noise-induced transitions, as well as noise-induced transport phenomena \cite{hanggi_2009,roberto_2013}.

\section{The model}
We consider a damped random-frequency harmonic oscillator whose
temporal evolution reads as \cite{gitterman_2005}
\begin{equation}\label{random_osc}
\frac{d^2 x}{dt^2} + \Gamma \frac{dx}{dt} + \omega_{0}^{2}\cor{1 +
\phi\pare{t}}x = 0,
\end{equation}
where $\Gamma$ is the damping coefficient, $\omega_{0}$ is the
average frequency of the oscillator and $\phi\pare{t}$ describes a
stochastic Gaussian process with zero average $\ave{\phi\pare{t}}
= 0$, and a specific autocorrelation function defined by
$\ave{\phi\pare{t}\phi\pare{t'}} = \kappa\pare{t-t'}$, where the
function $\kappa\pare{t-t'}$ defines the type of noise that is
considered. For instance, in the case of ideal white noise, the
autocorrelation function is defined as $
\ave{\phi\pare{t}\phi\pare{t'}}= 2D\delta\pare{t-t'}$, where $D$
denotes the intensity of the noise. A more realistic example is
colored noise, where the autocorrelation function writes
$\ave{\phi\pare{t}\phi\pare{t'}}=
(D/\tau_{c})\exp\pare{-\abs{t-t'}/\tau_{c}}$, with $\tau_{c}$
being the correlation time of the stochastic process
\cite{book_neuro}.

\begin{figure}[t!]\label{circuit}
\begin{center}
       \includegraphics[width=8.9cm]{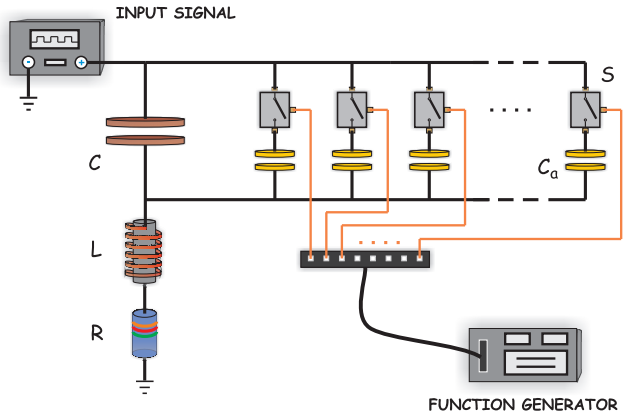}
\end{center}
\label{figure1} \caption{Scheme of the damped random-frequency
electrical oscillator consisting of a \emph{RLC} circuit with
central capacitance $C$, inductance $L$ and a parasitic resistance
$R$. Frequency fluctuations are achieved by switching on and off
individual capacitors $C_{a}$ by means of analog switches $S$ that
are driven by an arbitrary function generator.}
\end{figure}

Using the cumulant expansion described by Van Kampen
\cite{van_kampen_book}, Gitterman showed \cite{gitterman_book,gitterman_2005} that the equation for
$\ave{x}$, in the fast fluctuations regime, has the form (see Appendix for a detailed derivation)
\begin{equation}\label{ave_osc}
\begin{split}
\frac{d^2\ave{x}}{dt^2} + &\pare{\Gamma + \frac{\omega_{0}^{4}}{2\nu^{2}}\text{c}_{2}}\frac{d\ave{x}}{dt} \\
& + \omega_{0}^{2}\cor{1 -
\frac{\omega_{0}^{2}}{2\nu}\pare{\text{c}_{1} -
\frac{\Gamma}{2\nu}\text{c}_{2}}}\ave{x} = 0,
\end{split}
\end{equation}
where $\nu = \pare{\omega_{0}^{2} - \Gamma^{2}/4}^{1/2}$, and the
coefficients $\text{c}_{1}$ and $\text{c}_{2}$ are defined by
\begin{eqnarray}
\text{c}_{1} &=& \int_{0}^{\infty} \ave{\phi\pare{t}\phi\pare{t-\xi}}\sin\pare{2\omega_{0}\xi}d\xi, \\
\text{c}_{2} &=& \int_{0}^{\infty}
\ave{\phi\pare{t}\phi\pare{t-\xi}}\cor{1 -
\cos\pare{2\omega_{0}\xi}}d\xi.
\end{eqnarray}
Notice that, as pointed out in Ref. \cite{gitterman_book}, the existence of frequency fluctuations in Eq.
(\ref{random_osc}) introduces a noise-induced additional damping
and a noise-induced frequency shift to the average signal of the
oscillator.

\section{Experiment}

\subsection{The setup}

The experimental setup that allows us to introduce
random frequency fluctuations into a harmonic oscillator model is
the following. Firstly, note that one can construct a system
governed by Eq. (\ref{random_osc}) by making use of electrical
$RLC$ oscillators (where $R$ stands for resistance, $L$ for
inductance and $C$ for capacitance). In these systems, the charge
in the capacitor satisfies the same equation as Eq.
(\ref{random_osc}), where the coefficients $\Gamma$ and
$\omega_{0}$ are defined by \cite{berthet_2003}
\begin{eqnarray}
\Gamma &=& R/L, \\ \label{loss} \omega_{0} &=& \pare{L
C_{0}}^{-1/2}, \label{freq}
\end{eqnarray}
with $C_{0}$ denoting the average capacitance of the circuit. From
Eq. (\ref{freq}) one can see that fluctuations in the frequency of
the $RLC$ oscillator can be introduced by randomly switching the
values of the capacitance \cite{explanation}.

\begin{figure}[t!]\label{binom_dis}
\begin{center}
       \includegraphics[width=8.8cm]{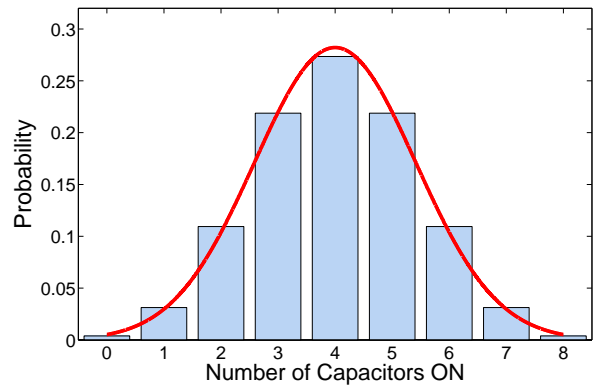}
\end{center}
\label{figure1} \caption{The probability that \emph{n}-capacitors
in the array are \emph{on} follows a binomial distribution, which
corresponds to a discretized Gaussian distribution with the same
average and variance. }
\end{figure}

Random switching of capacitance is performed in the following way:
An array of eight capacitors, each with equal capacitance $C_{a}$,
is connected in parallel to a central capacitor $C$ of a $RLC$
circuit. To produce uncorrelated random switching, the individual
capacitors are independently turned \emph{on/off} by means of
analog switches (NXP-74HC4066N quad bilateral switch), which are
driven by independent digital signals provided by an arbitrary
function generator (Signadyne digital I/O module
SD-PXE-DIO-H0001), as shown in Fig. 1. Because we are interested
in designing a Gaussian stochastic process, we program the
arbitrary function generator, so each capacitor has the same
probability to be \emph{on} or \emph{off}, in the same fashion as
in a coin-tossing event. It is easy to show that the probability
that $n$-capacitors in the array are \emph{on} satisfies a
binomial distribution given by
\begin{equation}\label{binom}
P\pare{n} = \binom{8}{n} \frac{1}{2^{8}},
\end{equation}
where $n=\{ 0, 1, 2, ... ,8\}$. This distribution is defined by a
mean value $\ave{n} = 4$ and a variance $\sigma^{2}_{b}=2$. Notice
that the binomial distribution described in Eq. (\ref{binom}) is a
discrete version of a Gaussian distribution with the same mean and
variance, as depicted in Fig. 2. It is important to remark that
due to the nonlinear relation between frequency and capacitance
[Eq. (\ref{freq})], when calculating the probability distribution
of frequency, a Gaussian distribution is obtained provided that
the condition $C_{a} \ll C$ is satisfied \cite{book_jacobs}.

\subsection{Implementation and Results}

To test the proposed scheme, we construct a $RLC$ circuit where
the central capacitance $C$ is provided by a 1 nF ceramic
capacitor, inductance $L$ by a 1.5 mH ferrite core inductor, and
resistance $R$ represents parasitic losses within the system. For
the random switching of capacitance, we have designed several
arrays using different ceramic capacitors with capacitance value
$C_{a} = \{ 4.7, 10, 18, 33, 47, 68, 100 \}$ pF. Notice from Fig.
2 that by changing the values of $C_{a}$ one can modify the
variance of the Gaussian distribution, which in turn modifies the
statistics of the noise in the system \cite{book_neuro}.

To guarantee that our system is well described by Eq. (\ref{ave_osc}), we make sure that frequency fluctuations are faster than the characteristic time evolution of the system, i.e., they satisfy the fast fluctuations condition \cite{gitterman_book}. To this end, the digital signals from the arbitrary function generator are set with a time rate
$\tau = 650$ ns, which is longer than the response time of the
analog switches ($400$ ns), and much shorter than the temporal
evolution window of the measured signal ($t = 100\; \mu$s).

\begin{figure}[t!]\label{circuit}
    \begin{center}
       \subfigure[]{\includegraphics[width=8.8cm]{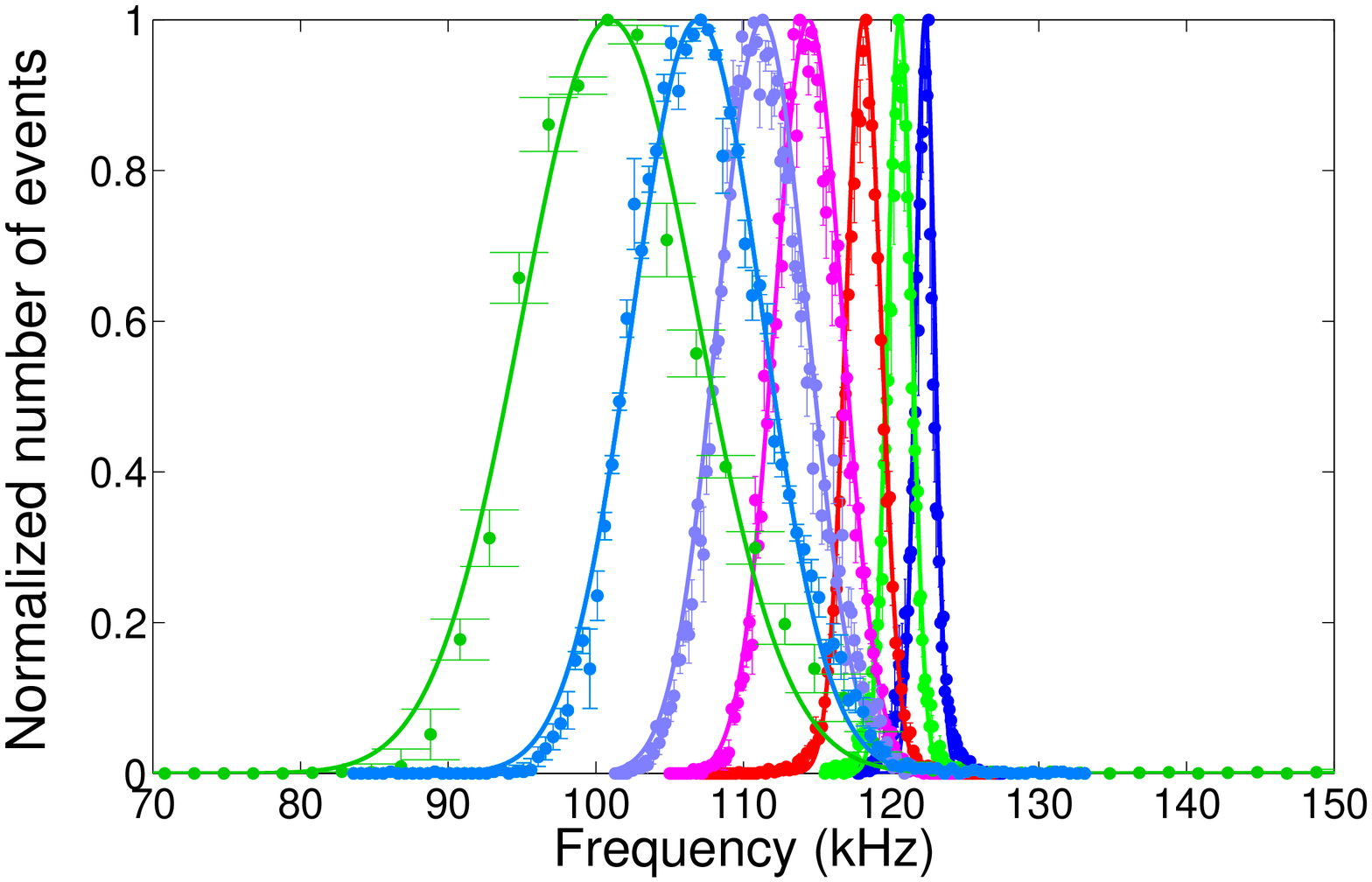}} \\
       \subfigure[]{\includegraphics[width=8.8cm]{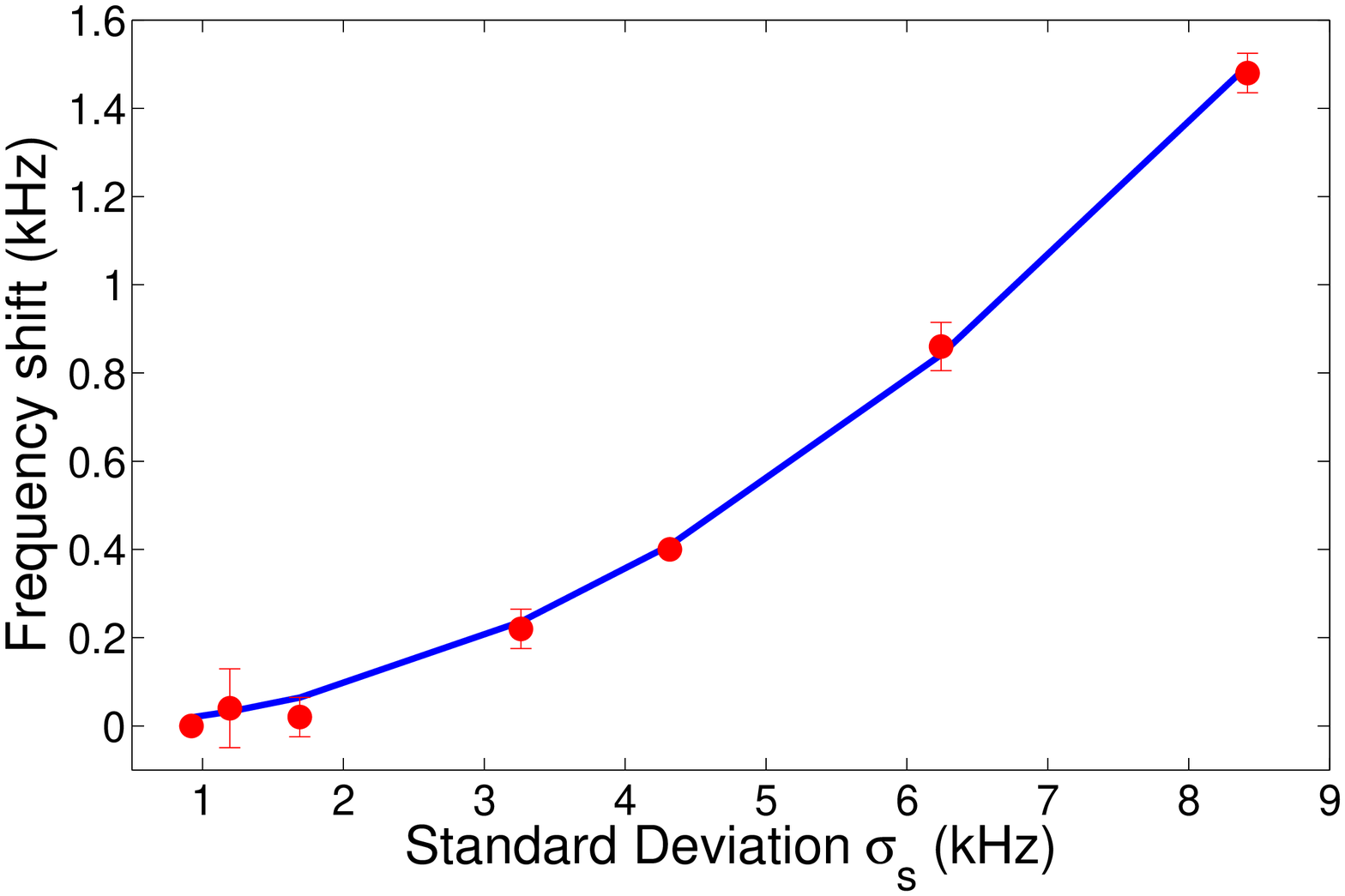}}
    \end{center}
\caption{(a) Signal frequency histograms using different capacitor
arrays. From left to right: $C_{a} = 100, 68, 47, 33, 18, 10, 4.7$
pF. Dotted line: Experimental data, Solid line: Gaussian fitting.
(b) Frequency shift as a function of the standard deviation
$\sigma_{s}$. Dotted line: Experiment, Solid line: Theory.}
\end{figure}

Using the system described above, we have performed the simulation
of Eq. (\ref{random_osc}). To this end, we keep the capacitor $C$
fixed and measure the averaged signal of the oscillator connected
to different capacitor arrays. Figure 3(a) shows the histograms of
the measured frequency in each case. Histograms are obtained from
$50000$ different realizations, and they are normalized to the
maximum number of events, where we define number of events as the
number of realizations that have the same value of frequency.
Notice that in all cases the probability distribution of the
frequency follows a Gaussian distribution whose variance
$\sigma_{s}^{2}$ depends strongly on the value of $C_{a}$ used in
the connected array. This implies that this scheme allows to
control the variance of the noise that is introduced in the
system. Moreover, notice that by changing the way in which capacitors are turned \emph{on}/\emph{off}, one can modify the frequency probability distribution. For instance, a dichotomous-like random frequency could be obtained by switching \emph{on} and \emph{off} all capacitors at the same time.

To compare the results obtained in Fig. 3(a) with the theoretical model, we have measured the frequency shift that
arises from the influence of frequency fluctuations, as predicted
by Eq. (\ref{ave_osc}). Figure 3(b) shows the frequency shift for
each capacitor array. We have made use of Eq. (\ref{ave_osc}), and
the relation \cite{book_neuro}: $D = \sigma^{2}\tau$, to find that
the driving noise of our system can be described by a
colored-noise-like autocorrelation function of the form
\begin{equation}
\ave{\phi\pare{t}\phi\pare{t'}}=
\frac{\sigma^{2}}{\omega_{0}^{2}}\exp\pare{-\frac{\abs{t-t'}}{\tau}},
\end{equation}
where the mean value of the frequency is computed with $C_{0} = C
+ 4C_{a}$, and the variance of the driving noise is $\sigma^{2} =
\pare{\eta\sigma_{s}}^{2}$, with $\eta = 3.4$. This relation
between both variances can be understood as a consequence of the
damping term in Eq. (\ref{random_osc}). The same effect can be
found, for instance, in the Ornstein-Uhlenbeck process, where the
resulting variance is proportional to the variance of the driving
noise due to the presence of a damping term
\cite{van_kampen_book}.

In general, when simulating noise-induced transport effects, one is interested in
keeping the mean frequency of each oscillator fixed while
increasing the strength of the noise
\cite{aspuru_2009,plenio_2008,roberto_2013}. This can be achieved
in our system by controlling the values of the central capacitance
$C$ and the time duration $\tau$ of the digital signals. Figure 4
shows the frequency histograms measured with different capacitor
arrays. Notice that by properly controlling the parameters of the
system, we are able to center all the probability distributions in
the same value of frequency $f_{0} \simeq 123$ kHz. This
demonstrates the flexibility of our system when modifying the
statistical properties of the environmental noise that interacts
with the oscillator. Parameters of the system used in each case
are summarized in Table 1.

\begin{figure}[t!]\label{binom_dis}
\begin{center}
       \includegraphics[width=8.8cm]{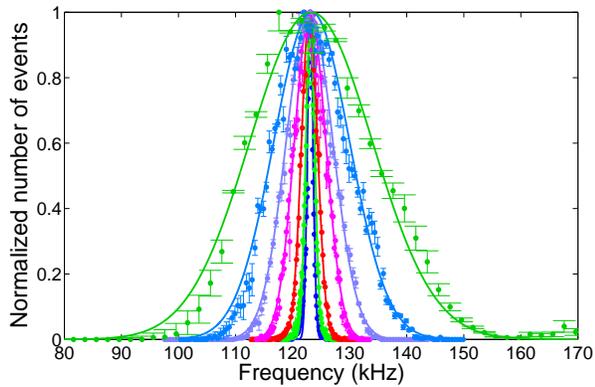}
\end{center}
\label{figure1} \caption{Frequency histograms for different
capacitor arrays centered in the same mean frequency $f_{0} \simeq
123$ kHz. Dotted line: Experimental data, Solid line: Gaussian
fitting.}
\end{figure}

\begin{table}[t]
\renewcommand{\arraystretch}{1.8}
\begin{tabular}{!{\vrule width 0.25mm} c  c !{\vrule width 0.25mm} !{\vrule width 0.25mm} c | c | c | c | c | c | c !{\vrule width 0.25mm}} 
\noalign{\hrule height 0.25mm}
$C_{a}$ (pF) && \hspace{2.2mm}4.7\hspace{2.2mm} & \hspace{2.2mm}10\hspace{2.2mm} & \hspace{2.2mm}18\hspace{2.2mm} & \hspace{2.2mm}33\hspace{2.2mm} & \hspace{2.2mm}47\hspace{2.2mm} & \hspace{2.2mm}68\hspace{2.2mm} & \hspace{2.2mm}100\hspace{2.2mm} \\
\hline 
$C$ (nF) && 1.120 & 1.090 & 1.053 & 0.978 & 0.933 & 0.840 & 0.355 \\
\hline 
$\tau$ (ns) && 650 & 650 & 750 & 780 & 800 & 720 & 650 \\
\noalign{\hrule height 0.25mm} 
\end{tabular}
\caption{Experimental parameters used to obtain the histograms
shown in Fig. 4.}
\end{table}

\section{Conclusions and Outlook}

In this paper, we have demonstrated a system that performs as a
tunable environment for classical electrical oscillators. We have
shown its operation by implementing the case of a damped
random-frequency oscillator, where a perfect agreement with the
theoretical model has been obtained. Finally, we have demonstrated
the degree of control that one can achieve with this system by
gradually modifying the variance of the frequency fluctuations,
while maintaining a fixed central frequency of oscillation, which
is of critical importance when simulating noise-induced energy transfer
mechanisms in different scenarios, such as in the case of energy transfer
in molecular aggregates.

The high degree of tunability and control of the proposed system can be further used to
design various types of noise with different probability
distributions. Moreover, it might allow us to study the transition
from Markovian to non-Markovian dynamics of open systems. The
results reported here represent an important step towards the experimental observation
of Gaussian and non-Gaussian noise-induced transitions, and noise-induced transport effects.

\acknowledgments

We thank Adam Vall\'{e}s, Luis Jos\'{e} Salazar Serrano, Yannick
de Icaza Astiz, Daniel Mitrani and Jos\'{e} Carlos Cifuentes for
valuable discussions. This work was supported by the projects
funded by the Government of Spain FIS2010-14831 and Severo Ochoa.
This work has also been partially supported by Fundacio Privada
Cellex Barcelona. J. Svozil\'{i}k acknowledges projects
CZ.1.07/2.3.00/30.0004 of M\v{S}MT of \v{C}R and PrF-2013-006 of
IGA UP Olomouc.

\appendix*
\section{Derivation of the averaged amplitude equation of a damped random-frequency harmonic oscillator}

Let us consider the equation for a damped random-frequency
harmonic oscillator
\begin{equation}\label{Eq:1}
\frac{d^2 x}{d t^2}+\Gamma\frac{d x}{d t}  +
\omega^2_0\left[1+\alpha\phi\left(t\right)\right]x=0,
\end{equation}
where $\Gamma$ is the damping coefficient, $\omega_{0}$ is the
average frequency of the oscillator, and $\phi\pare{t}$ is a
dimensionless stochastic variable with zero average
$\ave{\phi\pare{t}} = 0$, and an autocorrelation function
satisfying $\langle\phi(t_1)\phi(t_2)\rangle\rightarrow0$ for any
two time points $t_{1}$, $t_{2}$ such that $|t_2-t_1|>\tau_c$,
where $\tau_c$ is the correlation time of the stochastic process.
To simplify the derivation of Eq. (\ref{ave_osc}), and for
consistency with Ref. \cite{van_kampen_book}, we have included in
Eq. (\ref{Eq:1}) the parameter $\alpha$, which represents the
strength of the stochastic fluctuations.

In order to solve Eq. (\ref{Eq:1}), we first transform it into a
set of first-order differential equations
\begin{equation}\label{matrix_form}
\frac{d}{dt} X  = \pare{M_{\text{d}} + M_{\text{s}}} X,
\end{equation}
where
\begin{equation}
X = \left[ \begin{array}{c} x\pare{t}\\ \dot{x}\pare{t}
\end{array}\right],
\end{equation}
\begin{equation}
M_{\text{d}} = \left[\begin{array}{cc}
0 & 1 \\
-\omega_0^2 & -\Gamma
\end{array}
\right],
\end{equation}
\begin{equation}\label{Ms}
M_{\text{s}} = \left[\begin{array}{cc}
0 & 0 \\
-\alpha\omega^2_0\phi\left(t\right) & 0
\end{array}\right].
\end{equation}
Here, the matrices $M_{\text{d}}$ and $M_{\text{s}}$ represent the
deterministic and stochastic evolution of the oscillator,
respectively, and $\dot{x}\pare{t}$ stands for the time derivative
of the oscillator's amplitude $x$.

In the matrix representation, it is easy to show that the equation
for the deterministic evolution of the oscillator, i.e.
$\dot{X}_{\text{d}} = M_{\text{d}}X_{\text{d}}$, has the solution
\begin{equation}
X_{\text{d}}\pare{t} = U_{\text{d}}\pare{t} X_{\text{d}}\pare{0},
\end{equation}
where
\begin{widetext}
\begin{equation}\label{transform}
\begin{split}
U_{\text{d}}\pare{t} = \exp\pare{-\frac{\Gamma}{2}t}
\left[\begin{array}{cc} \cos\left(\nu
t\right)+\pare{\Gamma/2\nu}\sin\left(\nu t\right) &
\sin\left(\nu t\right)/\nu\\
- \omega_0^{2}\sin\left(\nu t\right)/\nu & \cos\left(\nu
t\right)-\pare{\Gamma/2\nu} \sin\left(\nu t\right)
\end{array}\right].
\end{split}
\end{equation}
\end{widetext}
Notice that the presence of damping in the harmonic oscillator
produces a frequency shift that is given by
$\nu=\sqrt{\omega_0^{2}-\Gamma^2/4}$.

Now, we make use of Eq. (\ref{transform}) to perform the
transformation
\begin{equation}\label{Eq:9}
X\pare{t} = U_{\text{d}}\pare{t}\tilde{X}\pare{t},
\end{equation}
which by substituting it into Eq. (\ref{matrix_form}) allows us to
write
\begin{equation}
\frac{d}{dt} \tilde{X} =\alpha
U_{\text{d}}\left(-t\right)M_{\text{s}}\left(t\right)U_{\text{d}}\left(t\right)
\tilde{X}\left(t\right). \label{Eq:10}
\end{equation}
Then, we iteratively solve Eq. (\ref{Eq:10}) to find that the
average of $\tilde{X}$ writes
\begin{equation}
\begin{split}
\langle\tilde{X}\pare{t}\rangle = & \langle\tilde{X}\pare{0}\rangle + \alpha^{2}\int_{0}^{t}dt_{1}\int_{0}^{\infty} dt_{2} \langle U_{\text{d}}\pare{-t_{1}}M_{\text{s}}\pare{t_{1}} \\
&\times
U_{\text{d}}\pare{t_{1}-t_{2}}M_{\text{s}}\pare{t_{2}}U_{\text{d}}\pare{t_{2}}\rangle\langle
\tilde{X}\pare{0}\rangle. \label{Eq:11}
\end{split}
\end{equation}
Notice that the linear term with $\alpha$ disappears since
$\langle M_{\text{s}}\pare{t}\rangle=0$. In writing Eq.
(\ref{Eq:11}), we have considered only the contributions up to
$\alpha^{2}$, which is an approximation that is valid as long as
the condition $\alpha\tau_c\ll1$ is satisfied. In addition, we
have assumed that the correlation time $\tau_{c}$ is much shorter
than the integration time, so we can take
$\langle\tilde{X}\pare{t}\rangle \rightarrow \langle
\tilde{X}\pare{0}\rangle$, and integrate to infinity, in the
second term of Eq. (\ref{Eq:11}).

We now perform the time derivative of Eq. (\ref{Eq:11}) to obtain
\begin{equation}
\begin{split}
\frac{d}{dt}\langle\tilde{X}\pare{t}\rangle = & \alpha^{2}\int_{0}^{\infty}d\xi \langle U_{\text{d}}\pare{-t}M_{\text{s}}\pare{t}U_{d}\pare{\xi}M_{\text{s}}\pare{t-\xi} \\
&\times U_{\text{d}}\pare{t-\xi}\rangle\langle \tilde{X}\pare{0}
\rangle,
\end{split}
\label{Eq:12}
\end{equation}
where the substitutions $\xi=t_1-t_2$, and $t = t_{1}$ are used.
Inverse transformation of Eq. (\ref{Eq:12}), by means of Eq.
(\ref{Eq:9}), then gives
\begin{equation}
\begin{split}
\frac{d}{dt}\langle X\pare{t} \rangle = & \left[ M_{\text{d}} + \alpha^{2}\int_{0}^{\infty}d\xi \langle M_{\text{s}}\pare{t}U_{\text{d}}\pare{\xi}\right. \\
&\times M_{\text{s}}\pare{t-\xi}U_{\text{d}}\pare{-\xi}\rangle
\bigg]\langle X\pare{t}\rangle.
\end{split}
\label{Eq:13}
\end{equation}
Using Eqs. (\ref{Ms}) and (\ref{transform}), we can readily find
that the expression inside the integral of Eq. (\ref{Eq:13})
writes
\begin{widetext}
\begin{equation}
\begin{split}
\langle M_{\text{s}}\pare{t}U_{\text{d}}\pare{\xi}M_{\text{s}}\pare{t-\xi}U_{\text{d}}\pare{-\xi}\rangle = & \omega_0^4\langle\phi\left(t\right)\phi\left(t-\xi\right)\rangle \\
& \times \left\{\begin{array}{cc}
0 & 0 \\
\cor{\sin\left(\nu \xi\right)/\nu}\left[\cos\left(\nu
\xi\right)-\pare{\Gamma/2\nu}\sin\left(\nu \xi\right)\right] &
-\sin^{2}\left(\nu \xi\right) / \nu^2
\end{array}\right\}.
\end{split}
\label{Eq:14}
\end{equation}
\end{widetext}

Finally, by substituting Eq. (\ref{Eq:14}) into Eq. (\ref{Eq:13}),
and transforming the set of two first-order equations to a single
second-order differential equation, Eq. (\ref{ave_osc}) is
obtained.


\begin{thebibliography}{XX}

\bibitem{hanggi_1998} L. Gammaitoni, P. H\"{a}nggi, P. Jung, and F. Marchesoni, Rev. Mod. Phys. \textbf{70}, 223 (1998).

\bibitem{horsthemke_book} W. Horsthemke and R. Lefever, \emph{Noise-Induced Transitions: Theory and Applications in Physics, Chemistry, and Biology} (Springer, Berlin, 1984).

\bibitem{parrondo_1994} C. Van den Broeck, J. M. R. Parrondo, and R. Toral, Phys. Rev. Lett. \textbf{73}, 3395 (1994).

\bibitem{aspuru_2009} P. Rebentrost, M. Mohseni, I. Kassal, S. Lloyd, and A. Aspuru-Guzik, New J. Phys. \textbf{11}, 033003 (2009).

\bibitem{plenio_2008} M. Plenio and S. Huelga, New J. Phys. \textbf{10}, 113019 (2008).

\bibitem{hanggi_2009} P. H\"{a}nggi and F. Marchesoni, Rev. Mod. Phys. \textbf{81}, 387 (2009).

\bibitem{roberto_2013} R. de J. Le\'on-Montiel and Juan P. Torres, Phys. Rev. Lett. \textbf{110}, 218101 (2013).

\bibitem{kawakubo_1978} T. Kawakubo, S. Kabashima, and Y. Tsuchiya, Progr. Theor. Phys. \textbf{64}, 150 (1978).

\bibitem{kabashima_1979} S. Kabashima and T. Kawakubo, Phys. Lett. A \textbf{70}, 375 (1979).

\bibitem{kabashima2_1979} S. Kabashima, S. Kogure, T. Kawakubo, and T. Okada, J. Appl. Phys. \textbf{50}, 6296 (1979).

\bibitem{berthet_2003} R. Berthet, A. Petrossian, S. Residori, B. Roman, and S. Fauve, Physica D \textbf{174}, 84 (2003).

\bibitem{residori_2001} S. Residori, R. Berthet, B. Roman, and S. Fauve, Phys. Rev. Lett. \textbf{88}, 024502 (2001).

\bibitem{zautkin_1983} V. V. Zautkin, B. I. Orel, and V. B. Cherepanov, Sov. Phys. JETP \textbf{58}, 414 (1983).

\bibitem{john_1999} T. John, R. Stannarius, and U. Behn, Phys. Rev. Lett. \textbf{83}, 749 (1999).

\bibitem{wio_2004} H. S. Wio and R. Toral, Physica D \textbf{193}, 161 (2004).

\bibitem{fuentes_2001} M. A. Fuentes, R. Toral, and H. S. Wio, Physica A \textbf{295}, 114 (2001).

\bibitem{castro_2001} F. J. Castro, M. N. Kuperman, M. Fuentes, and H. S. Wio, Phys. Rev. E \textbf{64}, 051105 (2001).

\bibitem{bouzat_2004} S. Bouzat and H. S. Wio, Eur. Phys. J. B \textbf{41}, 97 (2004).


\bibitem{gitterman_book} M. Gitterman, \emph{The Noisy Oscillator: The First Hundred Years, From Einstein Until Now} (World Scientific, Singapore, 2005).

\bibitem{graham1982} R. Graham, M. H\"{o}hnerbach, and A. Schenzle,
Phys. Rev. Lett. \textbf{48}, 1396 (1982).

\bibitem{ishimaru1999} A. Ishimaru, \emph{Wave Propagation and
Scattering in Random Media} (IEEE Press, Piscataway, NJ, 1997).

\bibitem{turelly1977} M. Turelli, Theoretical Population Biology
\textbf{12}, 140 (1977).

\bibitem{takayasu1997} H. Takayasu, A.-H. Sato, and M. Takayasu, Phys.
Rev. Lett. \textbf{79}, 966 (1997).

\bibitem{gitterman2013} M. Gitterman and D. A. Kessler, Phys. Rev. E \textbf{87}, 022137 (2013).

\bibitem{gitterman_2005} M. Gitterman, Physica A \textbf{352}, 309 (2005).

\bibitem{book_neuro} C. Laing and G. J. Lord, \emph{Stochastic Methods in Neuroscience} (Clarendon Press, Oxford, 2008)

\bibitem{van_kampen_book} N. G. Van Kampen, \emph{Stochastic Processes in Physics and Chemistry} (Elsevier, The Netherlands, 2007).

\bibitem{explanation} For the sake of simplicity, we have selected a random switching of the capacitance. However, one can always choose to randomly change the values of the inductance. By doing this, one adds more complexity to the system since the damping coefficient would randomly fluctuate as well.

\bibitem{book_jacobs} K. Jacobs, \emph{Stochastic Processes for Physicists: Understanding Noisy Systems} (Cambridge University Press, UK, 2010).

\end{thebibliography}
\end{document}